\documentclass[preprint,pre,floatfix,superscriptaddress]{revtex4}
\usepackage{graphicx}
\usepackage{AMSsymb}
\usepackage{epic,eepic}
\usepackage{graphics}
\usepackage{epsfig}
\usepackage{subfigure}

\newcommand{\uv}{{\bm u}}

\newcommand{\Lml}{L(t)}
\newcommand{\Kml}{K(t)}

\newcommand{\T}{T}

\newcommand{\dt}{{\Delta t}}
 
\newcommand{\pii}{{\partial_i}}

\newcommand{\be}{\begin{equation}}
\newcommand{\ee}{\end{equation}}
\newcommand{\bea}{\begin{eqnarray}}
\newcommand{\eea}{\end{eqnarray}}

\newcommand{\RT}{Rayleigh-Taylor}

\usepackage{amsmath}
\usepackage{amsfonts}
\usepackage{amssymb}
\usepackage{bm}
\usepackage{color}
\usepackage{graphicx}

\begin{document}
\title{High resolution numerical study of Rayleigh-Taylor turbulence
  using a thermal lattice Boltzmann scheme }

\author{L. Biferale} \affiliation{Department of Physics and INFN,
  University of Tor Vergata,\\ Via della Ricerca Scientifica 1, 00133
  Rome, Italy \\ and International Collaboration for Turbulence
  Research} \author{F. Mantovani} \affiliation{Deutsches
  Elektronen-Synchrotron, Platanenallee 6, 15738 Zeuthen, Germany}
\author{M. Sbragaglia} \affiliation{Department of Physics and INFN,
  University of Tor Vergata, \\ Via della Ricerca Scientifica 1, 00133
  Rome, Italy} \author{A. Scagliarini} \affiliation{Department of
  Physics and INFN, University of Tor Vergata,\\ Via della Ricerca
  Scientifica 1, 00133 Rome, Italy \\ and International Collaboration
  for Turbulence Research} \author{F. Toschi} \affiliation{Department
  of Physics and Department of Mathematics and Computer Science, and
  J.M. Burgerscentrum, Eindhoven University of Technology, 5600 MB
  Eindhoven, The Netherlands; INFN, Ferrara,\\ via G. Saragat 1, 44100
  Ferrara, Italy and International Collaboration for Turbulence
  Research} \author{R. Tripiccione} \affiliation{Dipartimento di
  Fisica, Universit\`a di Ferrara and INFN, Ferrara,\\ via G. Saragat
  1, 44100 Ferrara, Italy}

\begin{abstract}

  We present results of a high resolution numerical study of two dimensional
(2d) Rayleigh-Taylor turbulence using a recently proposed  thermal lattice
Boltzmann method (LBT). The goal of our study is both methodological and
physical.  We assess merits and limitations concerning  small-  and large-scale
resolution/accuracy  of the adopted integration scheme. We discuss
quantitatively  the requirements needed to keep the method stable and precise
enough to simulate stratified and unstratified flows driven by thermal active
fluctuations at  high Rayleigh and high Reynolds numbers. We present data with
spatial resolution up to $ 4096\times 10000$ grid points and Rayleigh number up
to $Ra \sim 10^{11}$.  The statistical quality of the data allows us to
investigate velocity and temperature fluctuations, {\it scale-by-scale},  over
roughly four  decades. We present a detailed quantitative analysis of scaling
laws in the viscous, inertial and integral range, supporting the existence of a
Bolgiano-like inertial scaling, as expected in 2d systems. We also discuss the
presence of small/large intermittent deviation to the scaling of
velocity/temperature fluctuations and the Rayleigh dependency of gradients
flatness. 

\end{abstract}

\maketitle
\section{Introduction}
\label{sec:introduction}
The \RT\ (RT) instability is present whenever we have the
superposition of a heavy fluid above a lighter one in a constant
acceleration field \cite{chandra}. Applications are numerous, from
inertial-confinement fusion \cite{rt2} to supernovae explosions
\cite{rt3} and many others \cite{rt.review}.  The RT instability has
been studied for decades, but it still presents several open problems
\cite{rt4}. It is important to control the initial and asymptotic
evolution of the mixing layer between the two miscible fluids; the
small-scale turbulent fluctuations, their anisotropic/isotropic ratio;
their dependency on the initial perturbation spectrum, on the geometry
of the containing volumes or on the physical dimensions of the
embedding space (see \cite{jot,boffi} for recent high resolution
numerical studies).  Concerning astrophysical and nuclear
applications, the two fluids evolve with strong compressible and/or
stratification effects, a situation which is difficult to investigate
either theoretically or numerically.  The set up studied in this paper
is two dimensional (2d) and the initial configuration slightly
different from what usually found in the literature: the spatial
temporal evolution of a single component fluid with a cold uniform
region in the top half and a hot uniform region on the bottom half
(see figure \ref{fig:rt} for details).  Such a situation is of
interest for convection in the atmosphere, ocean or even stars
interiors, where, masses of hot/cold fluid may be found in unstable
situations \cite{rt_atmosphere,rt_atmosphere2,rt_atmosphere3}.  The choice to focus
on a 2d geometry is motivated by different methodological, theoretical
and phenomenological challenges. First, concerning the method, 2d
geometries allow to push the numerics to unprecedented resolution -
here up to $4096 \times 10000$ grid points - with correspondingly high
Rayleigh/Reynolds numbers; this is an excellent testing ground for the lattice Boltzmann Thermal (LBT) scheme
\cite{POF.nostro,JFM.nostro} in fully developed situations, with
highly intermittent gradient statistics, and a well developed inertial
range of scales with power law distributions. We initially validate
the method against exact relationships originating from the
hydrodynamical Navier-Stokes-Fourier equations. Then, within the
limits settled by the validation steps, we show that the scheme
-albeit being only second order accurate- allows for quantitative
studies of hydrodynamical statistical fluctuations over a {\it four}
decades interval of scales.  This is, to the best of our knowledge,
the first time such a huge range of scales has ever been explored
using LBT codes for turbulent flows.  From the phenomenological point
of view, theoretical work \cite{chertkov,chertkov2} and pioneering
numerical simulations \cite{celani1} at smaller resolution tell us
that \RT\ dynamics in 2d displays Bolgiano statistics for velocity and
temperature fields, at least at scales small enough and far enough
from the edges of the mixing layer.  Bolgiano theory, at variance from
Kolmogorov theory \cite{Frisch}, predicts for typical inertial-range
velocity and temperature fluctuations on a generic inertial  scale, $R$, the
following  laws \be
\label{eq:bolgiano}
\delta_R T \sim \left( \frac{R}{\Lml} \right)^{1/5};\qquad \delta_R u \sim \Kml  \left( \frac{R}{\Lml} \right)^{3/5},
\ee
where $\Lml$ is a measure of the extension of the mixing layer at any given time
$t$ during the RT evolution, and $\Kml$ is the square root of the
total kinetic energy inside the mixing layer (see below for a precise
definition). These scaling properties tell us that
temperature/velocity is rougher/smoother than expected for Kolmogorov
scaling $\sim R^{1/3}$. This is due to the active role played by
buoyancy in the vertical momentum evolution, i.e. temperature becomes
a fully active scalar at all inertial  scales. 
This is in clear
contrast with the Kolmogorov like phenomenology expected
\cite{chertkov} and observed \cite{boffi,boffi2010} in three
dimensional (3d) cases. 2d \RT\ systems realize one of those cases
where the forcing mechanism --buoyancy-- overwhelms non-linear energy
transfer. This has also theoretical relevance, connected to the
universality of small-scale statistics in presence of multi-scale
forcing mechanisms \cite{pandit,bif_prl,doering,vassilicos,bif_njp} in
general, or to renormalization group approaches \cite{mazzino_rg}, in
particular. At variance with stochastic external multi-scale
forcing mechanisms, here the statistics of the buoyancy is directly
connected to the velocity field itself, opening the way for new
phenomena which we discuss in details later. Far from being
interesting only for theoretical reasons, Bolgiano scaling is believed
to characterize small scale velocity and temperature fluctuations in
3d Rayleigh-B\`enard convection close to the rigid boundaries, where
the viscous and thermal boundary layers merge with the bulk region
\cite{detlef_arfm}. In fact, thermo-hydrodynamical evolution in the
proximity of the boundaries is considered to be the key ingredient
driving the whole cell behavior \cite{Lohse}.

Here we will be mainly interested to small scale
properties, even though large scale evolution presents  many important open
issues, in particular for stratified flows. For example, we have recently shown
that RT evolution in the set-up of figure \ref{fig:rt} is stopped by the
adiabatic gradient in presence of a strongly stratified atmosphere
\cite{POF.nostro}. Investigation of small scale properties of such situation, as
well as the overshooting observed at the edge of the mixing layer is in progress
and will be reported elsewhere. 

All simulations are performed using an innovative LBT, proposed in
\cite{JFM.nostro} and already validated concerning large scale
properties on the same geometry here investigated
\cite{POF.nostro}. Stable, accurate and efficient discrete kinetic
methods describing simultaneous hydrodynamical evolution of momentum
and internal energy are notoriously difficult to achieve
\cite{succi,wolf}.  The main difficulties stem from the development of
subtle instabilities when the velocity increases locally. In recent
years, the situation has started to improve, as different attempts
have been made to describe active thermal modes within a fully
discretized Boltzmann approach
\cite{LL,Prasianakis,Sofonea,Gonnella,watari,Shan06,Philippi06,Nie08,Meng09}.

The advantages offered by LB codes are threefold. First, the
hydrodynamical manifold is described by the whole
Navier-Stokes-Fourier equations, with no need to rely on
incompressible and Boussinesq like approximations. Second, the method
is particularly efficient in dealing with complex bulk or boundary
physics, opening the way to incorporate either surface tension effects
or complex boundary conditions.  Last but not least, pressure
fluctuations are fully incorporated in the hydrodynamical evolution,
so we do not need to solve for Poisson equations; the method becomes
fully local in space, allowing for efficient implementations on
massively parallel machines, even if limited interconnection is
available.  Building on this point, our numerical results have been
obtained on the QPACE system, a massively parallel machine that uses
PowerXCell 8i processors connected by a toroidal network \cite{QPACE1,
  QPACE2}, following the lines of similar older attempts
\cite{lbeape}.

Results are as follows. In section \ref{sec:RT} we present the notation and
the main physical quantities that we study in this note, including a cursory
overview of RT large scale properties. In section \ref{sec:LBM} we briefly
summarize  the LBT method, we present the numerical details  and we discuss the
validation steps.  In section \ref{sec:smallscales} we present our results on
statistical fluctuations of temperature, velocity, temperature-fluxes and
buoyancy terms over the whole range of scales accessed by our numerics. We show
that velocity statistics is Bolgiano-like with  very small -if any- intermittent
corrections. We discuss the possible origin of these small anomalous corrections,
in relation with the corresponding  small intermittent fluctuations of the
buoyancy term, a new scenario for 2d turbulence. On the other hand, we show that
temperature fluctuations are strongly intermittent with high-order moments fully
dominated by hot/cold fronts. Such strong intermittency has a direct influence
also on the temperature flux statistics. Our resolution allows
us to address quantitatively and {\it scale-by-scale} the statistical properties
of all hydrodynamical fields; this analysis has not been accessible to
earlier 2d numerical studies \cite{celani1} and it is still not within reach in
the 3d case. 
Our concluding remarks (section \ref{sec:conclusions}) discuss
possible further development towards the study of (i) reactive \RT\
systems; (ii) strongly stratified systems; (iii)
multiphase/multi-component \RT\ or convection systems.

\begin{table*}
\begin{center}
\begin{tabular}{|c | c c c c c c c c c c c c c |}
  \hline & $At$ & $L_x$ & $L_z$ & $\nu$ & $k$ & $g$ & $T_{up}$ & $T_{down}$ & ${\tau}$ & $L_\gamma$  & $\eta(\tau)$ & $Ra_{max}$ & $N_{conf}$ \\
\hline run (A) & $0.05$ & 4096 & 10000 & 0.005 & 0.005 & $ 2 \times 10^{-5}$ & $0.95$ & $1.05$ & $6.4 \times 10^4$ & 10000 & 4.3  & $8 \times 10^{9}$  & 18 \\
run (B) &$0.05$& 4096 & 6000 &0.0025 & 0.0025 & $ 2.67 \times 10^{-5} $ & $0.95$ & $1.05$ & $ 5.5 \times 10^4$ & 7500 & 2.2 & $2 \times 10^{10}$ & 5 \\
run (C) &$0.05$& 4096 & 6000 &0.001 & 0.001 & $2.67 \times 10^{-5} $ & $0.95$ & $1.05$ & $ 5.5 \times 10^4$ & 7500 & 1.5 &  $1 \times 10^{11} $ & 23 \\
\hline
\end{tabular}
\caption{Parameters for the three types of RT runs. Atwood number  $At=(T_d-T_u)/(T_d+T_u)$; viscosity $\nu$; thermal diffusivity $k$; gravity $g$; temperature in the upper half region $T_u$;  temperature in the lower half region $T_d$; normalization time ${\tau}=\sqrt{L_x/(g\;At)}$; adiabatic lenght corresponding to the adiabatic gradient $L_\gamma = \Delta T/\gamma$; dissipative scale calculated at $t=\tau$, $\eta(\tau)$; Maximum Rayleigh number $Ra_{max}$; number of independent RT evolution $N_{conf}$.}
\label{table:param} 
\end{center}
\end{table*}

\section{\RT\ systems}
\label{sec:RT}
The spatio-temporal evolution of a stratified compressible flow, in a
external gravity field, $g >0$, is ruled by the Navier-Stokes-Fourier
equations (double indexes are meant summed upon) : \be
\begin{cases} 
  \label{1}
  D_t \rho = - \rho \pii u_i  \\
  \rho D_t u_i = - \pii P - \rho g\delta_{i,z} + \mu \partial_{jj}
  u_i  \\
  \rho c_p D_t T - D_t P = \chi \partial_{ii} T ,
\end{cases}
\ee 
\noindent where $D_t$ is the material derivative, $\mu,\chi$ the molecular
viscosity and thermal conductivity, $c_p$ the specific heat at
constant pressure and $\rho, T, P, \uv$ are density, temperature,
pressure and velocity field. Under the assumption that compressibility
and stratification are small (the situation addressed in this note)
and that fluid parameters depend weakly on the local thermodynamic
fields, one can expand pressure around its hydrostatic value $ P = P_0
+ p$, with $\partial_z P_0 = -g \rho$ and $p\ll P_0$, and perform a
small Mach number expansion \cite{spiegel1,spiegel3}: \be
\begin{cases}
\label{1.a}
D_t u_i = - \frac{\pii p}{\rho} + \frac{g \theta}{T_m} \delta_{i,z}   + \nu \partial_{jj} u_i  \\
D_t T -u_z \gamma = k \partial_{ii} T .
\end{cases}
\ee 
In this approximation, only temperature fluctuations $\theta$ force
the system; we have introduced the mean temperature $T_m$, kinematic
viscosity $\nu = \mu/\rho$, thermal diffusivity $k = \chi/(c_p \rho)$
and adiabatic gradient for an ideal gas, $\gamma = g/c_p$. The small
Mach expansion and small stratification decouple the pressure from the
internal energy equation, i.e. $p$ in (\ref{1.a}) is just a Lagrange
multiplier used to enforce $\partial_i u_i =0$ everywhere. As we will
show in the next section, the LBT algorithm we are going to use is
meant to reproduce the set of equations (\ref{1}) and (\ref{1.a}) in
the corresponding limit.

\begin{figure}
\begin{center}
 \advance\leftskip-0.15cm
  \includegraphics[scale=0.36]{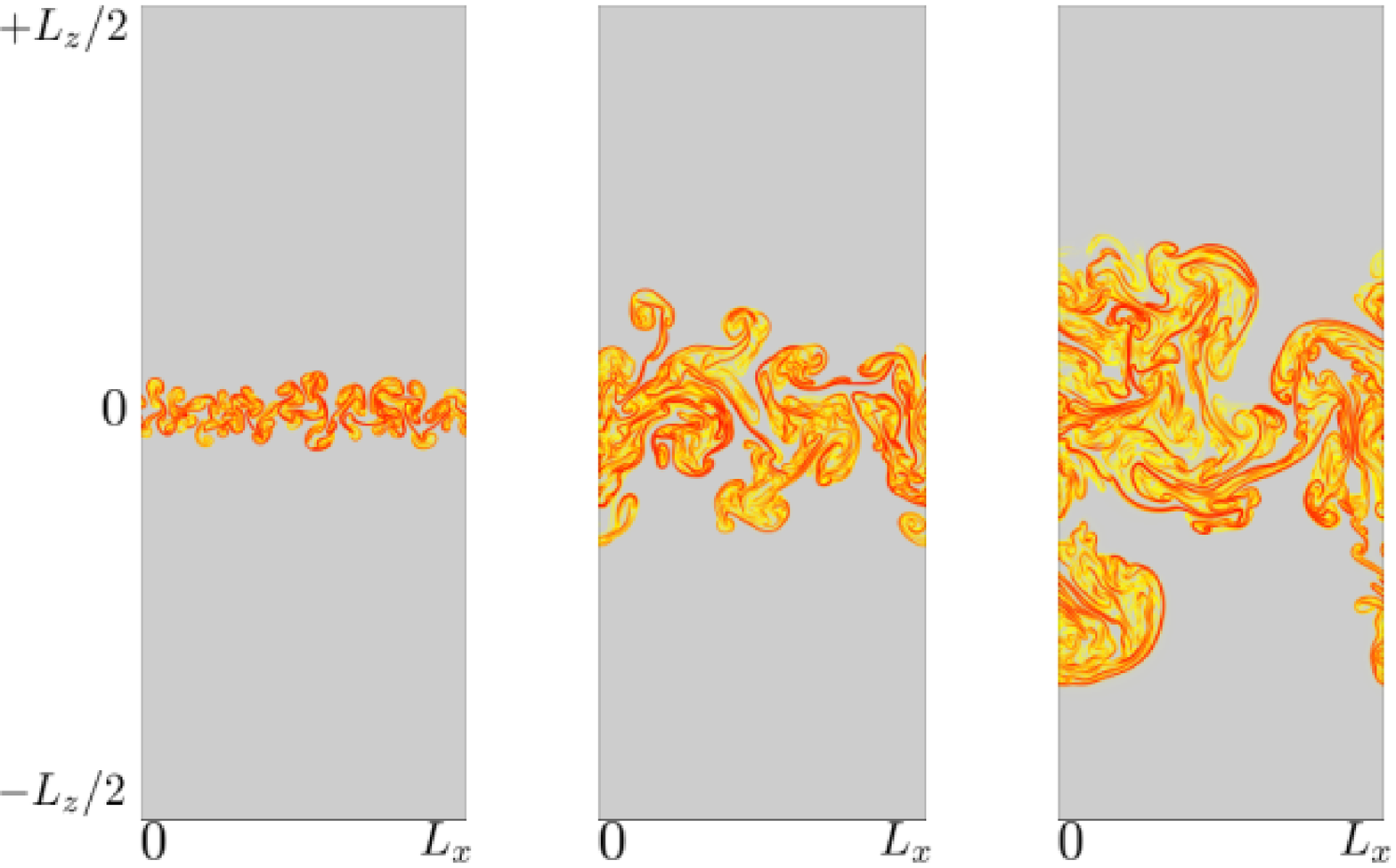}
 \advance\leftskip-0.55cm
  \includegraphics[scale=0.7]{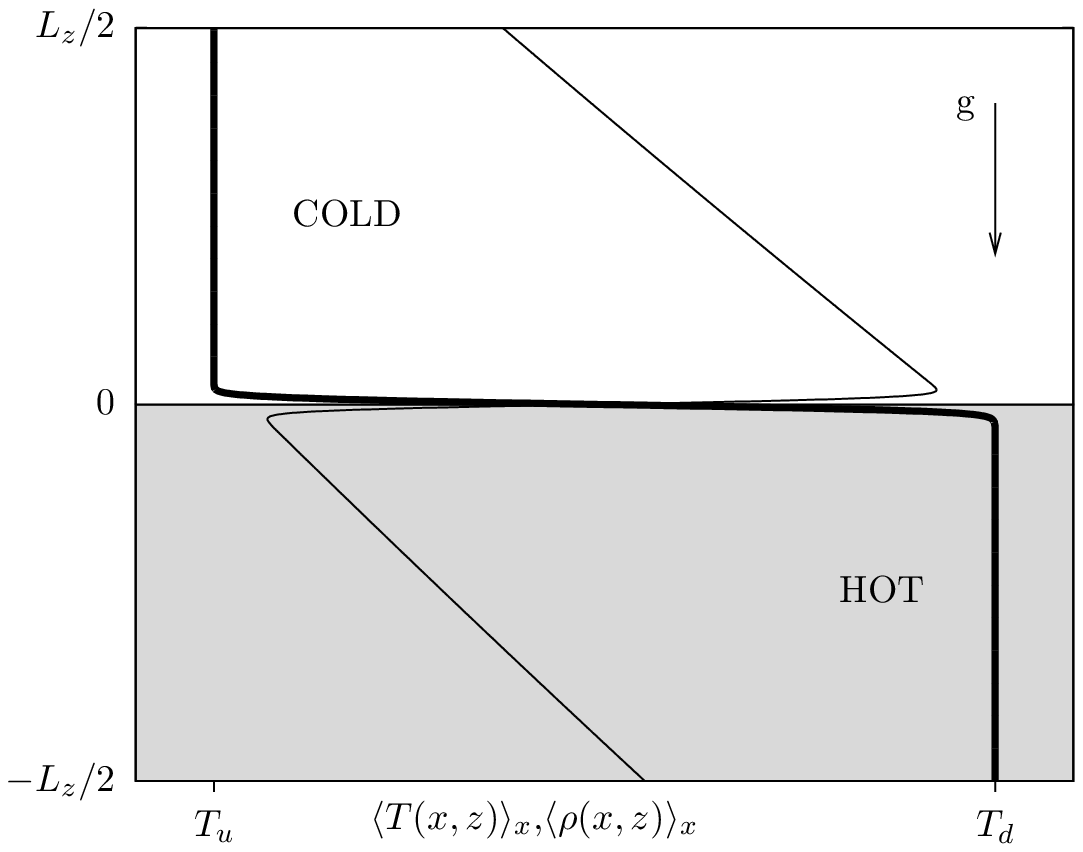} 
  \caption{ Bottom: Initial configuration for the  stratified \RT\ systems.  Temperature in the upper half is chosen constant $ T_0(z) = T_{up}$ while density follow an hydrostatic profile, $\rho_0(z) = \rho_{up} \exp(-g(z-z_c)/T_{up})$, with $z_c$ the central location in the box. In the lower half we have:   $T_0(z) = T_{down}$; and $ \rho_0(z) = \rho_{down} \exp(-g(z-z_c)/T_{down}).$ To be at equilibrium, we require to have the same pressure at the interface,  $ \rho_{up} T_{up} = \rho_{down} T_{down}.$  The temperature jump at the interface is smoothed by a $tanh$ profile with a width of the order of 10 grid points. The bold and  tiny solid lines represent the temperature  and density profiles respectively. Top: Snapshot of the RT evolution at three times $t =( 0.5,1,4) \tau$. }
\label{fig:rt}
\end{center}
\end{figure}

\begin{figure}
\begin{center}
\advance\leftskip-0.55cm
  \includegraphics[scale=0.7]{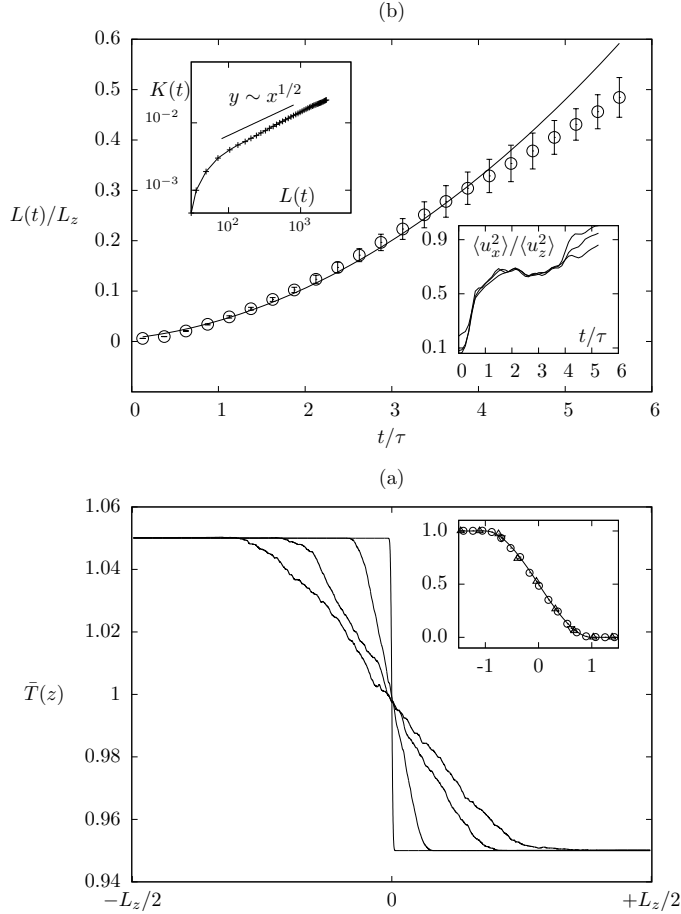}
  \caption{(a) Mean temperature profile at four different times during  the RT evolution. In the inset we show the rescaling according to the instantaneous mixing layer length $\Lml$, $(\bar T (z/\Lml,t)-T_{up})/(T_{down}-T_{up})$. The profile rescales perfectly and in agreement with the cubic shape predicted by a Prandtl mixing length theory \cite{boffi2010} (solid line). (b) Evolution of mixing layer, $\Lml$, with superposed the best parabolic fit (solid line), using the self-similar prediction 
(\ref{eq:t2}).
 Lower inset: ratio between horizontal $\langle u_x^2\rangle$  and vertical $\langle u_z^2\rangle$ kinetic energy, calculated in the whole, half or one quarter of the mixing layer: a transition around $\tau \sim 4$ 
is clearly visible.  Despite of this slowing down, the relative scaling of total kinetic energy with respect to the mixing layer length satisfies 
the scaling (\ref{eq:t2}). This is shown in the upper inset where we have $K(t) \sim L^{1/2}(t)$.}
\label{fig:L}
\end{center}
\end{figure}

If the adiabatic gradient is negligible, $\gamma \sim 0$, it is well
known that starting from an unstable initial condition as depicted in
figure \ref{fig:rt}, any small perturbation will lead to a turbulent
mixture between the hot and cold regions, expanding along the vertical
$z$-direction.  Concerning large scale quantities, a huge amount of
earlier work (e.g. see Ref. \cite{rt4}) has focused on the extimation
of the growth rate of the mixing layer extension, $\Lml$, and of the
total turbulent kinetic energy, $K^2(t) = \frac{0.5}{ L_x \Lml} \int dx
dz { u^2}$, produced by the conversion of the initial potential
energy. Using dimensional analysis and self-similar assumptions
\cite{cook,cook2} one predicts:
%
\be \Lml \sim \alpha (t+t_0)^2, \qquad \Kml \sim \beta
t; \label{eq:t2} \ee where $t_0$ is the typical time needed for the
system to reach a fully non-linear evolution.  The values of the
coefficients, $\alpha,\beta$, have been extensively studied both in 2d
and 3d
\cite{rt4,cook,celani1,rt.temp,cabot,young,p1,p2,POF.nostro}. They
depend on the definition of $\Lml$, typically taken either as the
region where the mean temperature profile, averaged over the
horizontal direction, $\bar T(z) = \frac{1}{L_x} \int dx T(x,z,t) $,
is within a given range, for example: $\bar T(z) \in 0.95[ T_{up}:
T_{down}]$, or as an integral property over the whole temperature
distribution: \be
\label{eq:lcook}
\Lml = \frac{1}{L_x} \int dx dz\, \Theta \left[
  \frac{T(x,z,t)-T_{up}}{T_{down}-T_{up}} \right], \ee with $ \Theta[x]
= 2 x;\;0 \le x \le 1/2$ and $ \Theta[x] = 2\,(1-x);\; 1/2 \le x \le 1
$.  Using the estimate (\ref{eq:t2}) one may predict the whole profile
evolution, adopting either simple constant eddy viscosity models or
more refined Prandtl mixing length theory \cite{boffetta.prl.2010}. In
figure \ref{fig:L} we show the growth rate of the mixing layer,
kinetic energy and the temporal evolution of the temperature profile,
as an example of typical evolutions of large scale quantities in our
numerics. The agreement with the expected phenomenology is very
satisfactory. Notice a systematic small deviation at large times. This
deviation is probably due to a transition induced by the evolving
aspect ratio.  When the aspect ratio becomes order one, important
horizontal fluctuations develop in the system, preventing an efficient
conversion of potential energy in vertical kinetic energy (inset of
the same figure).

In this paper, on the other hand, we focus on small scales quantities,
i.e. velocity, temperature and fluxes statistics scale-by-scale. In
particular we focus on the following set of structure functions,
based on moments of order $p$ of velocity, temperature or mixed
increments: \be
\begin{cases}
  \label{eq:sf}
  S^{(p)}_{\theta}(R,t) = \langle |\delta_R \theta|^p \rangle\\
  S^{(p)}_{u_i}(R,t) = \langle |\delta_R u_i|^p \rangle; \qquad i=x,z \\
  S^{(p)}_{B}(R,t) = \langle |\delta_R \theta| |\delta_R u_z|^p \rangle\\
  S^{(p)}_{F}(R,t) = \langle [(\delta_R \theta)^2 |\delta_R u_z|]^{p/3} \rangle,\\
\end{cases}
\ee

where we define the increment of a generic hydrodynamical field,
$A(x,z,t)$ as $\delta_R A = A(x+R,z,t)-A(x,z,t)$ and the average $$
\langle (\cdot) \rangle =\frac{1}{L_x \times L_z}\int_0^{L_x} dx
\int^{\tilde L_z/2}_{- \tilde L_z/2} dz\, (\cdot)$$ is performed on
the whole horizontal direction and on a given vertical range inside
the mixing layer. In order to minimize non homogeneous contributions,
we typically restrict the vertical extension of the averaging region
to $\tilde L_z = \frac{1}{2} \Lml$, with $\Lml$ estimated according to
the volume average (\ref{eq:lcook}). Moreover, in the correlation
functions defined above, we only show the results for spatial
increments along the fully homogeneous horizontal direction, $\hat
x$. Subscript $(B)$ and $(F)$ in the third and fourth row of
(\ref{eq:sf}) denote the correlation functions driving the time
evolution of the $p$-th moment of velocity increments (the buoyancy
forcing term) and of the temperature flux, respectively.  Chertkov in
\cite{chertkov} developed a coherent phenomenology for small-scales 2d
\RT\ systems, on the reasonable assumptions that (i) the mixing layer
evolution is adiabatically slow compared to small scales fluctuations;
(ii) the amount of kinetic energy dissipation at small scales is
negligible (absence of direct energy cascade in 2d turbulence); (iii)
temperature is efficiently dissipated at small scales (direct
temperature cascade). These three ingredients lead to a unique
possible dimensional prediction, the Bolgiano scaling
(\ref{eq:bolgiano}). In particular, one expects in the inertial range:

\be
\begin{cases}
\label{eq:sfinertial}
S^{(p)}_{\theta}(R,t) \sim (\frac{R}{\Lml})^{\zeta_\theta(p)} \\
S^{(p)}_{u_x,u_z}(R,t) \sim K^{p}(t) (\frac{R}{\Lml})^{\zeta_u(p)} \\
S^{(p)}_{B}(R,t) \sim K^{p}(t) (\frac{R}{\Lml})^{\zeta_{B}(p)} \qquad  \eta(t) \ll R \ll \Lml\\
S^{(p)}_{F}(R,t) \sim K^{p/3}(t) (\frac{R}{\Lml})^{\zeta_{F}(p)}, \\
\end{cases}
\ee
while in the viscous range: 
\be
\label{eq:sfviscous}
\begin{cases}
S^{(p)}_{\theta}(R,t) \sim (\frac{\eta(t)}{\Lml})^{\zeta_\theta(p)}(\frac{R}{\eta(t)})^p \\
S^{(p)}_{u_x,u_z}(R,t) \sim K^{p}(t) (\frac{\eta(t)}{\Lml})^{\zeta_u(p)} (\frac{R}{\eta(t)})^p  \\
S^{(p)}_{B}(R,t) \sim K^{p}(t) (\frac{\eta(t)}{\Lml})^{\zeta_{B}(p)} (\frac{R}{\eta(t)})^p    \qquad  R \ll \eta(t) \\
S^{(p)}_{F}(R,t) \sim K^{p/3}(t) (\frac{\eta(t)}{\Lml})^{\zeta_{F}(p)}(\frac{R}{\eta(t)})^p ,
\end{cases}
\ee
with 
\be
\label{eq:zpBolgiano}
\zeta_\theta(p) = \frac{p}{5}, \qquad \zeta_u(p) = \frac{3}{5}p; \ee
and \be \zeta_{B}(p) = (\zeta_\theta(1) + \zeta_u(p)), \hspace{.1in}
\zeta_{F}(p) = (\zeta_\theta(2)+\zeta_u(1)) \frac{p}{3}.  \ee
Moreover, according to 2d Bolgiano scaling, the dissipative scale
increases with time, as $\eta(t) \sim t^{1/8}$. The two expressions
(\ref{eq:sfinertial}-\ref{eq:sfviscous}) for inertial and viscous
ranges are such that they match at the viscous scale, $\eta(t)$.  The
presence of a non stationary evolution makes the problem particularly
interesting. The above phenomenology has been already investigated
numerically in \cite{celani1}, where a good
agreement with Bolgiano scaling for low order velocity structure
functions and a departure from Bolgiano dimensional scaling for
temperature structure functions were measured,
 for the first time. On one hand, the
results presented in \cite{celani1} clearly indicates the validity of
Chertkov's phenomenology, plus the extra complexity of anomalous
intermittent corrections to the temperature field. On the other hand,
due to limited spatial resolution, the authors of \cite{celani1} could
not assess statistical properties in a quantitative way {\it
  scale-by-scale} because they had scaling over only about a decade. Our data add
to the above discussion a detailed investigations of inertial, viscous
and integral range properties covering all together around 4
decades. We confirm and measure the presence of large anomalous
corrections to the temperature scaling:
$$
\zeta_\theta(p) = p/5 +  \Delta_\theta(p).
$$
We also show that our data cannot exclude the presence of small
deviations from Bolgiano scaling also for velocity field, a novel
observation, never reported before and somehow surprising for 2d
turbulence:
$$
\zeta_u(p) = 3p/5 + \Delta_u(p).
$$

 \section{Numerical method \& validation steps} \label{sec:LBM}

\subsection{The thermal lattice Boltzmann algorithm}
In this section we recall the essential features of the computational lattice
Boltzmann method employed in the numerical simulations. A complete analysis,
along with extensive validation steps, can be found in
\cite{POF.nostro,JFM.nostro}. The thermal-kinetic description of a compressible
gas/fluid with variable density $\rho$, local velocity ${\bm u }$, internal
energy ${\cal K}$, and subject to a local body force density ${\bm g}$, is given
by the following equations: 
\be 
\label{eq:1}
\begin{cases}
\partial_t \rho + \partial_i (\rho u_i) = 0 \\
\partial_t (\rho u_k) + \partial_i (P_{ik}) = \rho g_k \\
\partial_t {\cal K} + \frac{1}{2} \partial_i q_i = \rho g_i u_i ,
\end{cases}
\ee
where $P_{ik}$ and $q_i$ are the momentum and energy fluxes, still unclosed at
this level of description. A recent paper \cite{JFM.nostro} has
shown that it is possible to recover exactly  equations (\ref{eq:1}), starting
from a suitable discrete version of the Boltzmann equations with self consistent
local equilibria. The reference scheme is summarized by the following equation
set
\be\label{LBM}
f_{l}({\bm x}+ {\bm c}_{l} \dt,t+\dt) - f_{l}({\bm x},t)=-\frac{\dt}{\tau_{LB}}\left(f_{l}({\bm x},t) - f_l^{(eq)}({\bm x},t) \right),
\ee 
where $f_l({\bm x},t)$ represents a probability density function to find
a particle at space-time location $({\bm x},t)$ whose velocity ${\bm c}_l$ 
belongs
to a discrete set \cite{Philippi06,Shan06}. The lhs of equation (\ref{LBM})
stands for the streaming step of such probability whereas the rhs represents the
relaxation towards  local Maxwellian distribution function ${f}_{l}^{(eq)}$ with
characteristic time $\tau_{LB}$. 

\begin{figure}
\begin{center}
  \includegraphics[scale=0.3]{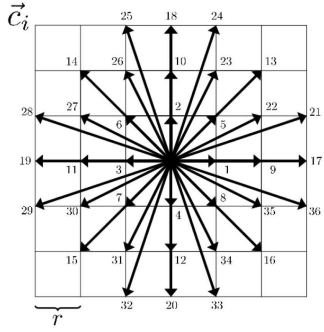}
  \caption{Scheme of the discrete set of velocities, $r$ is the lattice constant whose value is $r \approx 1.1969$ \cite{Philippi06,Shan06}. To recover the correct degree of isotropy for tensors describing thermal fluxes, one needs at least $37$ speeds in 2d and $105$ speeds in 3d. A smaller set of discrete velocities can be used if off-grid vectors are allowed \cite{Surmas09}.}
\label{fig:37}
\end{center}
\end{figure}
The macroscopic fields (density, momentum and temperature) are defined
in terms of the lattice Boltzmann populations: \be \rho = \sum_l
f_l;\;\; \rho {\bm u} = \sum_l {\bm c}_l f_l;\;\; D \rho \T = \sum_l
\left|{\bm c}_l - {\bm u}\right|^2 f_l, \ee with $D$ the space
dimensionality. The novelty of the algorithm here employed stems from
the form of the equilibrium distribution function.  Here, it directly
depends on the coarse grained variables plus a shift from the local
body force term: \be {f}_{l}^{(eq)}={f}_{l}^{(eq)}\left(\rho,{\bm
    u}+\tau_{LB} {\bm g},T+\frac{\tau_{LB}(\Delta
    t-\tau_{LB})}{D}g^2\right).  \ee The detailed structure of this
equilibrium distribution function can be found in
\cite{POF.nostro,Shan06,Philippi06}. Lattice discretization also
induces corrections terms in the macroscopic evolution of averaged
quantities: both momentum and temperature must be renormalized by
discretization effects in order to recover the correct hydrodynamical
description from the discretized lattice Boltzmann variables. The
first correction to momentum is given by a pre- and post-collisional
average \cite{Buick00,Guo02,POF.nostro}:
$$ {\bm u}^{(H)}={\bm u}+\frac{\Delta t}{2} {\bm g} \label{eq:shift_lattice}$$ 
and the first non-trivial correction to the temperature field by \cite{JFM.nostro}:  
$$ {\T}^{(H)} = \T + \frac{(\dt)^2g^2}{4 D }. \label{eq:thydro}$$  
Using these ``renormalized'' hydrodynamical fields it is possible to
recover, using a Chapman-Enskog expansion
\cite{POF.nostro,JFM.nostro}, the standard thermo-hydrodynamical
equations for a compressible fluid with energy conservation. Such
procedure, applied to the kinetic equations (\ref{eq:1}), sets the
fluxes $P_{ik}$ and $q_i$ equal to their hydrodynamical counterpart
describing advection, dissipation and diffusion. In two dimensions
($D=2$), the resulting equations for the hydrodynamical fields are
those given in equations (\ref{1}) (for the explicit calculation see
\cite{POF.nostro}).
\subsection{Details of the numerical simulations} \label{sec:sim}

\begin{figure}
\begin{center}
\includegraphics[scale=0.7]{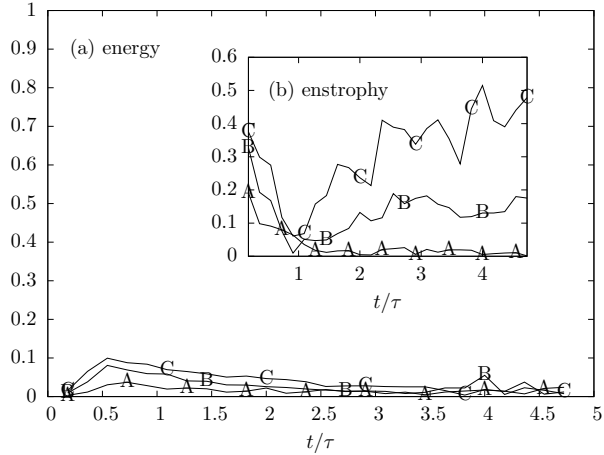}
\caption{ Large and small scales validation of the LBT scheme for the three sets of runs (A)-(C) in table I. (a) Difference between the lhs and the rhs  of the energy equation in (\ref{eq:validation}), normalized with the buoyancy term. (b) The same of (a) but for the enstrophy equation. 
\label{fig:fig.precision}
}
\end{center}
\end{figure}

\begin{figure}
\begin{center}
  \includegraphics[scale=0.7]{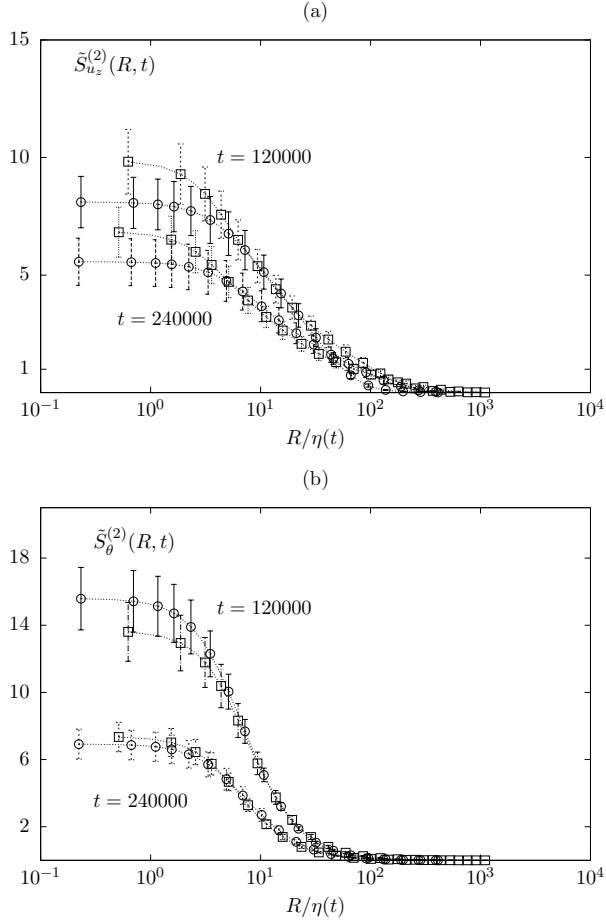}
  \caption{
Normalized effective gradients for runs (A) ($\circ$) and (C)  ($\square$) at
two different times along the RT evolution: $t = (120000,240000)$ in LBT units. }
\label{fig:isotropia}
\end{center}
\end{figure}

We use a 2d LBT algorithm, with 37 population fields  (the so called D2Q37
model), moving in the directions shown in figure \ref{fig:37}.  We have run on
the QPACE Supercomputer \cite{QPACE1, QPACE2}, a novel massively parallel
computer, powered by IBM PowerXCell 8i processors (an enhanced version of the
Cell processor) that supports our algorithm very efficiently \cite{lele1}. 
Three different sets of runs  have been performed (parameters are
summarized in table  \ref{table:param}) at varying accuracy: (A) a fully
resolved high resolution simulation, up to $4096 \times 10000$
collocation points with kinematic viscosity and thermal conductivity large
enough to ensure optimal  resolution of velocity and temperature fields even for
large order statistics; (B)  a less resolved high resolution 
simulation, up to $ 4096 \times  6000 $ collocation points, with small scale
transport parameters a factor 2 smaller than in  case (A); (C) a even less
resolved case with the same resolution of (B) and viscosity a factor 5 smaller
than (A).   Runs (B) and (C) make the Rayleigh and
Reynolds numbers as large as possible, 
even though the statistical properties for sub-viscous scales
will not be as accurate as for set (A).  The
remarkable result that we are able to present is that the LBT method
is able
to reproduce large scale and inertial range physics correctly even in those
cases (e.g. runs (B) and (C)), where very small scales are not resolved
correctly.
A systematic way to validate the accuracy of the method and its
convergence towards the hydrodynamical manifold of the kinetic
equations is to benchmark numerical results against exact
relationships coming from the hydrodynamical Navier-Stokes equations
of motions. For example, large and small scale accuracy can be checked
via the equations for the kinetic energy and the enstrophy of the
systems:

\be
\label{eq:validation}
\begin{cases}
\partial_t \frac{1}{2}\langle u^2 \rangle_V = -\epsilon_\nu + g \langle \theta u_z \rangle_V\\
\partial_t \frac{1}{2}\langle  w^2 \rangle_V = -\epsilon_\omega + g \langle \partial_x \theta w \rangle_V,\\
\end{cases}
\ee 
where the two dissipative terms are $\epsilon_\nu= \nu \langle (\partial_i u_j)^2 \rangle_V$ and $\epsilon_\nu= \nu \langle w^2 \rangle_V$,
and with $\langle (\cdot) \rangle_V$ we mean the average over the
whole volume. These two exact relations probe large and small scales,
respectively. In figure \ref{fig:fig.precision} we show the
percentage difference between left hand side and right hand side
normalized with the buoyancy term, for the three set of runs of table
\ref{table:param}.
Gradients of each field  have been calculated either as a centered difference of the hydrodynamical variable, or using the lattice definition
$$
\partial_i A({\bm x})  \approx  \sum_l  w_l c^i_{l} A({\bm x}+{\bm c}_l \Delta t)
$$
with $w_l$ suitable weights \cite{POF.nostro} and ${\bm c}_l$ the
lattice velocities (see figure \ref{fig:37}); we find that the second
choice gives better agreement.  While the energy balance equation is
well verified within a few percent for all resolutions, the enstrophy
balance for run (B) and (C) is not satisfactory. As a result, gradient
statistics will be measured only using data from run (A). The next
question concerns the range of scales at which accuracy becomes
acceptable also for runs (B) and (C).
 This can be monitored by
plotting a sort of normalized ``effective gradient'' at different
scales. In figure \ref{fig:isotropia} we show for temperature and
vertical velocity the quantities: $\tilde S^{(2)}_{u_z}(R,t) =
({\Lml}/{\eta(t)})^{\zeta_u(2)}S^{(2)}_{u_z}(R,t)/(\Kml R/\eta(t))^2$
and $\tilde
S^{(2)}_{\theta}(R,t)=({\Lml}/{\eta(t)})^{\zeta_\theta(2)}S^{(2)}_{\theta}(R,t)/(R/
\eta(t))^2$ at different times during the RT evolution. Clearly, even
though run (C) does not resolve gradients correctly, i.e. the curves
do not reach a well developed plateaux for small scales, they
superpose well with the well resolved run (A) as soon as $R \sim 5
\eta(t)$.  This result is important and
makes us confident that the LBT numerics is quantitatively accurate
even when small scales are not perfectly smooth, i.e. the method provides
for a sort of implicit large eddy simulation (ILES) with an effective
sub-grid dissipation.
\section{Small Scales Statistics}
\label{sec:smallscales}

As the mixing layer evolves, the effective Rayleigh number,
characterizing the thermal instability inside the layer, grows. In the
presence of stratification the expression for the Rayleigh number is not
unique. It is possible to introduce a $z$-dependent Rayleigh number \cite{spiegel1}: 
\be
\label{eq:Ra}
Ra(z,t) = \frac{(g/\bar T(z))L^4(t)(\frac{\Delta T}{\Lml} -\gamma)}{(k/\bar \rho(z)c_p)(\nu/  \bar \rho(z))}
\ee
where  the notation $\overline{(\cdot)}$  indicates
averages over the horizontal direction. We follow here a  common procedure defining  a  Rayleigh number based on the middle plane, i.e. $Ra(t)  = Ra(z=0,t)$. Notice
that the presence of stratification appears also through the adiabatic
term $\gamma$, i.e. any RT mixing of the kind here studied will be
stopped sooner or later  once an adiabatic atmosphere is reached. For
the case when the adiabatic term is not important, $\Delta T/\Lml \gg
\gamma$, the {\it ultimate scaling regime} predicted by Kraichnan is
expected. In this regime there is a
relationship between the normalized heat flux and the
Rayleigh number \cite{kraichnan,celani1,boffi,boffi2010}:
$$
Nu \sim Ra^{1/2}
$$
where the Nusselt number ($Nu$) is defined as the total heat flux inside the
mixing layer normalized with its conducting value: $Nu = \langle
\theta u_z\rangle/(k \Delta T/\Lml)$. 
\begin{figure}
\begin{center}
  \includegraphics[scale=0.7]{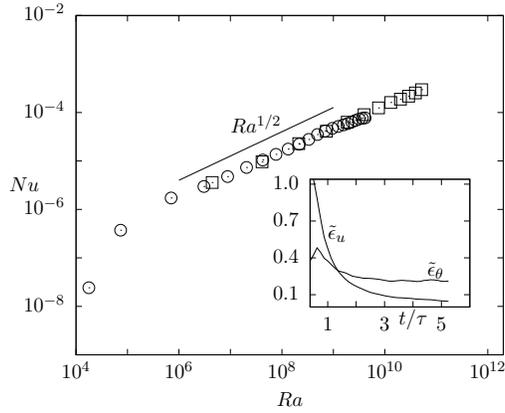}
  \caption{Nusselt  $vs$ Rayleigh, data from run (A) ($\circ$) and (C) ($\square$).  Inset: Dimensionless  dissipative anomalies, $\tilde \epsilon_\nu$ and $\tilde \epsilon_\theta$, at changing time during RT evolution.}
\label{fig:dissipation}
\end{center}
\end{figure}
Other important output parameters for the system are kinetic energy
dissipation and thermal dissipation,
$\epsilon_\nu,\epsilon_\theta$. In 2d we expect that the normalized
$\tilde \epsilon_\nu = \epsilon_\nu/(K^3(t)/\Lml)$ and $\tilde
\epsilon_\theta = \epsilon_\theta/((\Delta T)^2\Kml/\Lml)$ vanish and
go to a constant for large Rayleigh (Reynolds), respectively.  This is
tantamount to predict the existence of a direct cascade of temperature
fluctuations and the absence of a kinetic energy dissipation
anomaly. The monotonic increase of Rayleigh during the mixing
layer evolution allows for a check of the previous predictions. In
figure \ref{fig:dissipation} we show both the Nusselt vs. Rayleigh
law, confirming for more than 4 decades the observation of the {\it
  ultimate regime} and the behaviour of $\epsilon_\nu$ and
$\epsilon_\theta$ during the RT evolution. We observe the tendency
towards a constant non vanishing dissipative anomaly for temperature
fluctuations, while kinetic energy is becoming smaller and smaller at
increasing Rayleigh (Reynolds), as expected.

\subsection{Scale-by-scale statistics}
\begin{figure}
\begin{center}
  \includegraphics[scale=0.7]{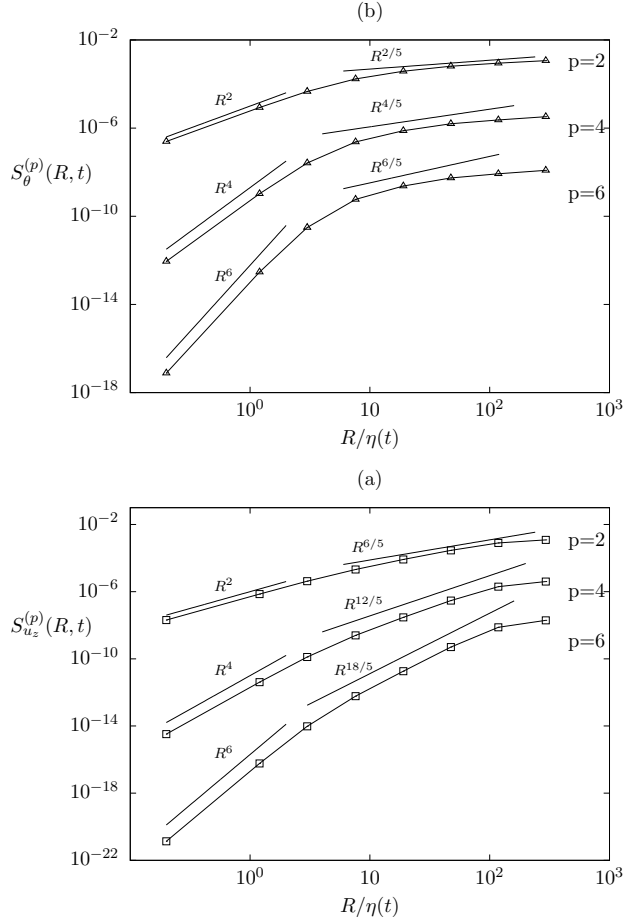}
  \caption{ Log-log plot of velocity (a) and temperature (b)scaling for $p=2,4,6$ at a late time during RT evolution ($\tau \sim 5$ and data from run (A)). We also plot the corresponding Bolgiano and viscous  scaling.   }
\label{fig:sf}
\end{center}
\end{figure}

\begin{figure}
\begin{center}
  \includegraphics[scale=0.7]{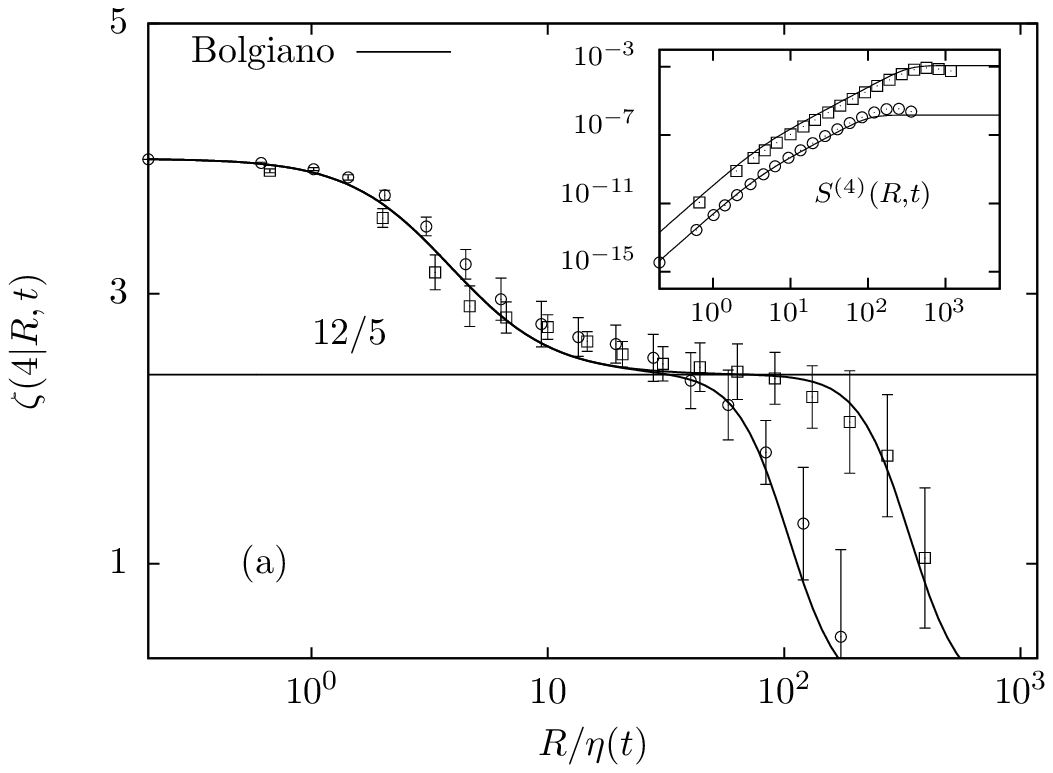}
  \includegraphics[scale=0.7]{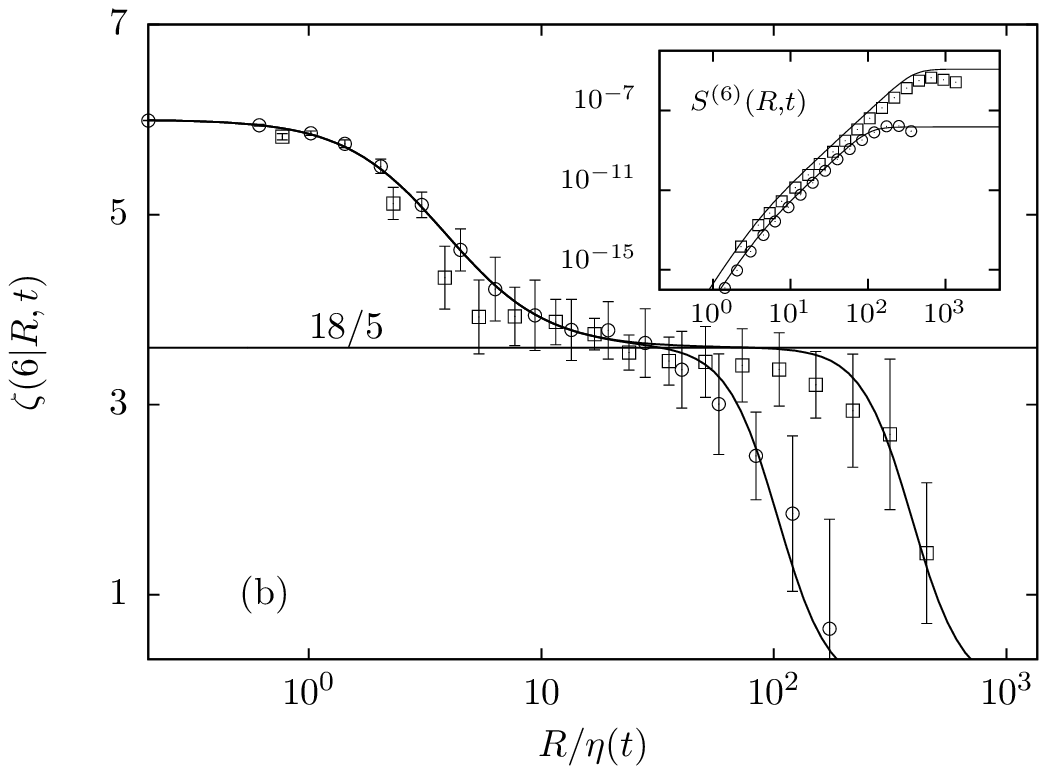}
  \caption{
Local scaling properties for velocity structure functions, $\zeta_{u_z}(p|R,t)$ for run (A) ($\circ$) at $t = 4 \tau$ and (C) ($\square$) at $t = 4 \tau$. We show the case with $p=4$ (panel a) and $p=6$ (panel b). Solid lines correspond to the {\it local scaling exponents} as predicted from  (\ref{eq:batchelor2}) using the Bolgiano dimensional scaling (\ref{eq:zpBolgiano}), also drawn as an horizontal line of value  $12/5$ and $18/5$ respectively. Error bars are calculates out of the scattering between the $N_{conf}$ different RT evolutions for each run. Insets: structure functions, $S_{u_z}^{(p)}(R,t)$, for $p=4$ and $p=6$ and the two runs (A) and (C) (same symbols). The solid line is the  parametrization (\ref{eq:batchelor2}).}
\label{fig:ls.fit.v4+6}
\end{center}
\end{figure}

\begin{figure}
\begin{center}
  \includegraphics[scale=0.7]{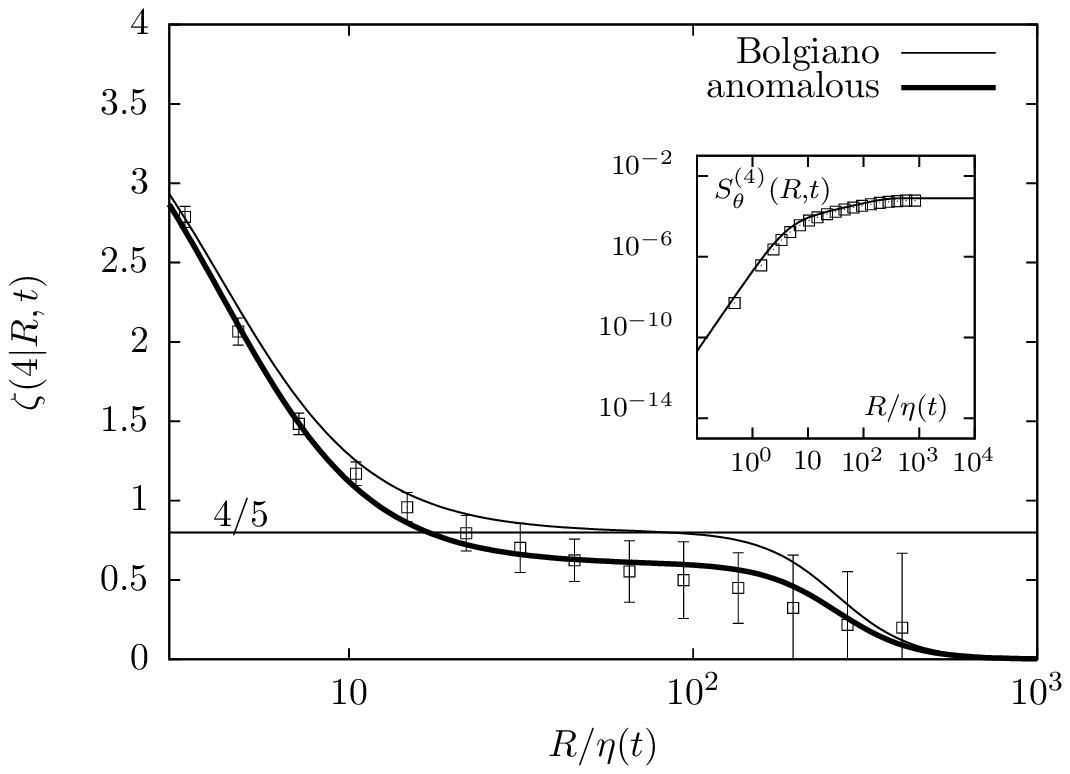}
  \includegraphics[scale=0.7]{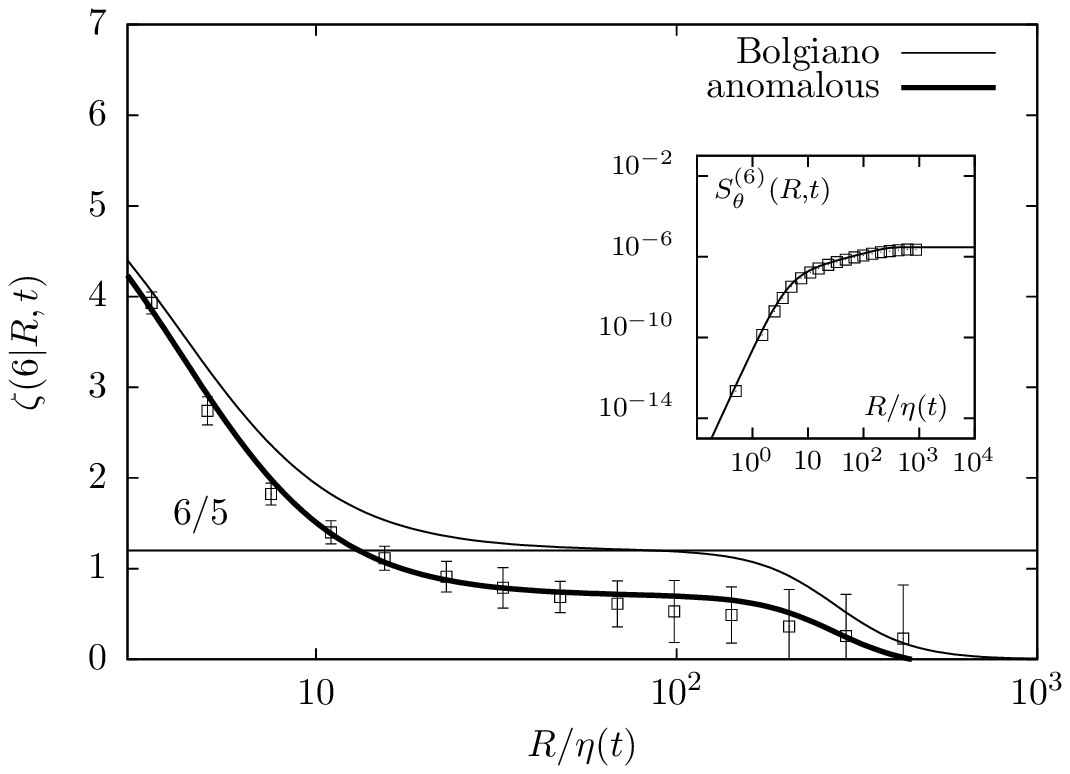}
  \includegraphics[scale=0.7]{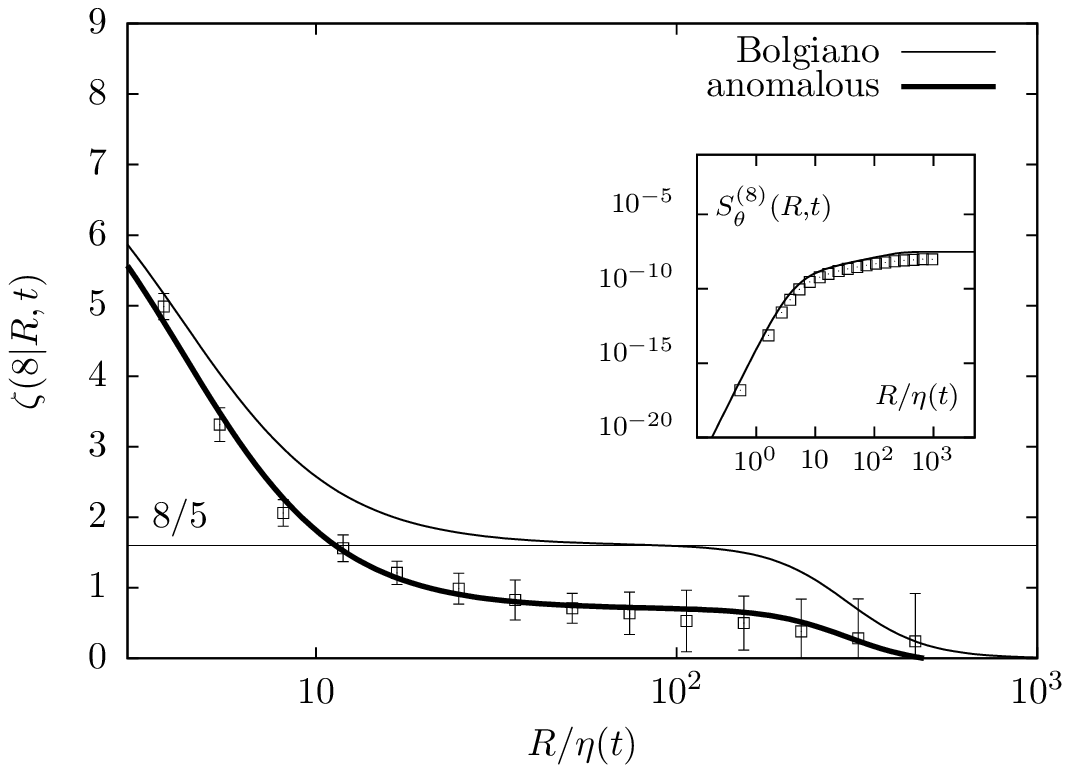}
  \caption{  
Same data as in figure \ref{fig:ls.fit.v4+6} but for temperature scaling. We show  $p=4$,   $p=6$ and   $p=8$ (panels from top to bottom). 
The solid line corresponds to the parametrization (\ref{eq:batchelor2}) using the dimensional Bolgiano scaling exponents. The thick solid line is the same parametrization but with anomalous scaling exponents. 
Already for $p=4$ and more importantly for $p=6$ and $p=8$ the LSE for the parametrization (\ref{eq:batchelor2}) with dimensional Bolgiano scaling  $\zeta(p)= p/5$  does not fit the numerical data. As a guide to the eyes, we also show the Bolgiano inertial range values as horizontal lines in each panel. The curves supporting  anomalous scaling  are obtained with the following correction to the exponents:$\Delta_\theta(4) = -0.2 $, $\Delta_\theta(6) = -0.5$ and $\Delta_\theta(8) = -0.7$. Insets: structure functions with superposed the Batchelor parametrisation with  anomalous inertial exponents (solid line).}
\label{fig:ls.fit.t4+6+8}
\end{center}
\end{figure}

In figure \ref{fig:sf} we show a log-log plot of
$S^{(p)}_\theta(R,t)$ and  $S^{(p)}_{u_z}(R,t)$,
for different orders and different times. We also superpose the
inertial  range scaling predicted by the dimensional
Bolgiano prediction. Even though on a log-log scaling
the global overall agreement between data and dimensional Bolgiano
scaling is not bad, important deviations can be seen both at the
crossover between viscous and inertial range, $R \sim \eta(t)$
 and around the integral scale, $R\sim \Lml$. Let us first investigate
 the viscous-inertial cross-over. There, typical velocity fluctuations
 have to go from a smooth differentiable behaviour $\delta_R u \sim
 R$ to Bolgiano scaling $\delta_R u \sim R^{3/5}$. The jump in the
 scaling property is therefore not too large, and one must expect
 important sub-leading contributions 
 well inside the inertial range coming from the viscous scaling.
Such sub-leading term may spoil
 scaling properties even at high Rayleigh values. Differently, for
 temperature, the jump in the scaling properties 
 from viscous to inertial is large (from $R^p$ to $R^{p/5}$). Sub-leading
 terms cannot play any role. Anyhow, such a big change in the scaling
 properties cannot happen in a too short range of scales: the
 interval of increments with neither a pure viscous nor a pure
 inertial scaling should be large in this case.\\ 
A ``conveniently simple transition function'' to encompass both viscous and inertial range scaling
 in a single fitting expression is given by the Batchelor
 parametrization \cite{batchelor,meneveau,sreeni,prl_2008,jfm_2010},
 which for a generic structure function of order $p$ reads:
\begin{equation}
\label{eq:batchelor1}
F^{(p)}(R,t) = C_p \frac{R^p}{(R^2 + A_p \eta^2(t))^{\frac{p-\zeta(p)}{2}}},
\end{equation}
where $C_p$ is a suitable dimensional normalization parameter and
$A_p$ is a dimensionless parameter taking into account some small
possible dependency of the viscous cutoff on the order of the
correlation function \cite{schumacher,yakhot,bb,bif}.  The above
expression is the simplest way to glue smoothly a differential
behaviour in the viscous range, $ \sim R^p$, for $R \ll \eta(t)$ with
a rough scaling, $\sim R^{\zeta(p)}$, in the inertial range, $\eta(t)
\ll R$. We need also to match the inertial-integral layer, $R \sim
\Lml$, where structure functions start to saturate because all
hydrodynamical fields decorrelate for $R \gg \Lml$. It is easy to
generalize the Batchelor parametrization to also encompass such a
region of scales, reaching a global phenomenological description of
structure functions valid for all scales:
\begin{equation}
\label{eq:batchelor2}
F^{(p)}(R,t) = C_p \frac{R^p}{(R^2 +
  A_p \eta^2(t))^{\frac{p-\zeta(p)}{2}}} \, (R^a + B\, L^{\alpha}(t))^{-\zeta(p)/a},
\end{equation}
where in the above expression,  the crossover around  $R \sim \Lml$ is fixed by the
parameter $a$ and $B$ (in the following always chosen $a=4$, $B=1$). The
potentialities of the parametrization (\ref{eq:batchelor2}) cannot be
appreciated on log-log plots:
 a detailed scale-by-scale analysis of structure
functions behaviour is needed. \\
A scale-by-scale analysis can be obtained by looking at the so-called
{\it Local Scaling Exponents} (LSE), i.e. the log derivatives of any
structure function:
\be
\label{eq:lse}
\zeta(p|R,t) = \frac{d \,\log (F^{(p)}(R,t)) }{ d \, \log (R)}.  \ee
Whenever we have a pure power law behaviour, the output must be a
constant as a function of the separation scale, $R$, $\zeta(p|R,t)
\sim \zeta(p,t)$. The advantage to measure (\ref{eq:lse}) stems from
the possibility to follow also the cross-over between viscous and
inertial range and between inertial and integral range,
scale-by-scale, hence the name.  In figure \ref{fig:ls.fit.v4+6} we
show for $p=4,6$ the velocity structure functions against the
Batchelor parametrization (\ref{eq:batchelor2}) for two different
Rayleigh numbers. 
In the body of the figure, we plot the LSE for
$S^{(p)}_{u_z}(R,t)$ from our data and superposed with the
corresponding expression coming from the parametrization
(\ref{eq:batchelor2}), where we have used the Bolgiano value
$\zeta_u(p) = \frac{3}{5}p$. The agreement is strikingly good;
considering together all data at different resolutions, we are able to
reproduce the viscous, inertial and integral scale behaviour over 4
decades of scaling range. The agreement between the Bolgiano
dimensional prediction and the velocity scaling is very accurate
within error bars. Notice that the use of LSE with respect to log-log
scaling as depicted in the inset of the same figure allows to move the
discussion from global fit over many orders of magnitude (for the
latter) to a scale-by-scale fit of ${\cal O}(1)$ quantities (for the
former). Moving to temperature scaling, the scenario changes. In
figure \ref{fig:ls.fit.t4+6+8} we show the same as figure
\ref{fig:ls.fit.v4+6} but for temperature and up to $p=8$. Here, the
agreement with the Batchelor parametrization with the Bolgiano
dimensional scaling for temperature $\zeta_\theta(p)=p/5$, is less
good, almost acceptable for low order moments, but definitely not on top of
the numerical data for high order moments. In order to achieve a good
fit, on the whole range of scale, one needs to introduce anomalous
corrections to the exponents $\zeta_\theta(p) = p/5 +
\Delta_\theta(p)$ used in the Batchelor formula. In the same figure we
show indeed how the use of $\Delta_\theta(4)= -0.2$,
$\Delta_\theta(6)= -0.5$, $\Delta_\theta(8)= -0.8$ gives a much better
agreement between numerical data and the phenomenological
parametrization formula (\ref{eq:batchelor2}). This is, in our view, a
very clean demonstration of the existence of anomalous scaling for
temperature fluctuations in 2d RT. The values measured for
$\Delta_\theta(p)$ are in agreement with the one presented in
\cite{celani1}. Result on temperature scaling are summarized in table (\ref{tab:temp}).
\begin{table}
\begin{center}
\begin{tabular}{|c | c c c |}
  \hline  $\zeta_\theta(p)$ & Bolgiano & Ref.[15]& here \\  
   \hline p=4 & 0.8  & 0.6 & 0.6 $\pm$ 0.06 \\
   \hline p=6 & 1.2  & 0.7 & 0.7 $\pm$ 0.07 \\
   \hline p=8 & 1.6  &  ---  & 0.8 $\pm$ 0.1 \\
\hline
\end{tabular}
\caption{Summary of temperature scaling exponents using the best fit obtained by the parametrization (\ref{eq:batchelor2}), using for $\eta(t)$ and $\Lml$ the actual 
values measured on the data}
\label{tab:temp} 
\end{center}
\end{table}

\begin{figure}
\begin{center}
  \includegraphics[scale=0.7]{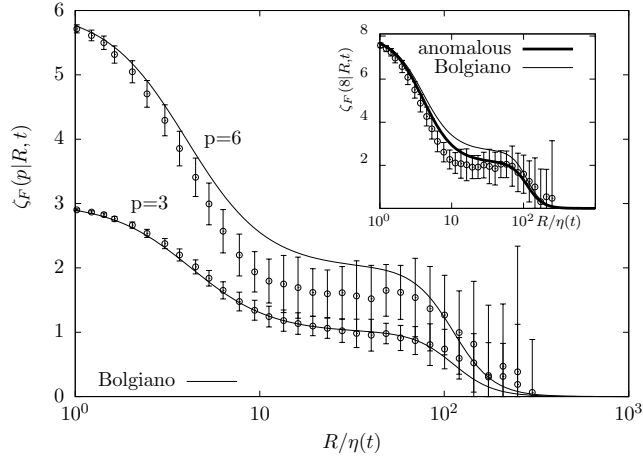}
  \caption{ Run (A). Local Scaling Exponent for temperature flux moments, 
$S_{F}^{(p)}(R,t)$, with $p=3,6$ and $p=8$ (inset). The solid thin line for $p=3,6$ (main plot) corresponds to the dimensional Bolgiano estimate  (\ref{eq:batchelor2}) with $\zeta(p) = (\zeta_\theta(2) + \zeta_{u_z}(2))p/3$. In the inset we show fit wioth both dimensional Bolgiano and anomalous estimates.  Notice
the better agreement with the anomalous case}
\label{fig:ls.fitvt2}
\end{center}
\end{figure}

\begin{figure}
\begin{center}
  \includegraphics[scale=0.7]{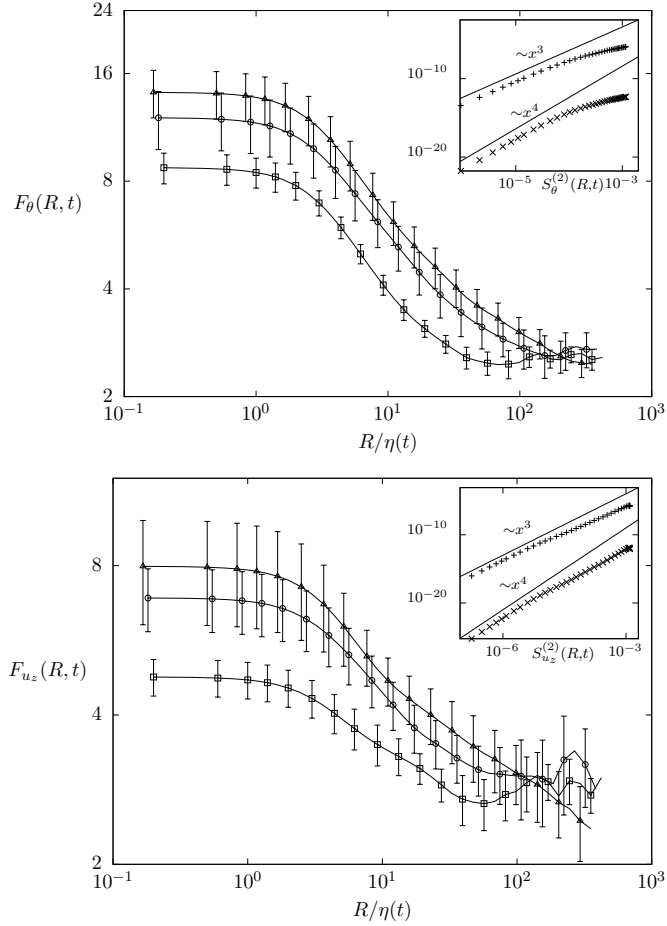}
  \caption{Velocity (bottom)  and temperature (top) flatness at three different times during RT evolution $t = (2,3,5)\tau$ in run (A). Insets: ESS plot for structure functions of order $p=6$ ($+$) and $p=8$ ($\times$)  versus structure function with $p=2$. The solid line corresponds to the dimensional scaling .}
\label{fig:ess}
\end{center}
\end{figure}

An important feature of RT in 2d, is the active role played by
buoyancy at all scales, as witnessed by the Bolgiano phenomenology.
The interesting point here is that, as the buoyancy is driven by
temperature fluctuations, the forcing mechanism in the momentum
equations is given by a non self-similar --intermittent--
field. Navier-Stokes equations forced with power law forcing, have
attracted the attention in the past both for application of the
renormalization group \cite{mazzino_rg} and for issues concerning
small-scales universality, i.e. understanding how strong must be the
forcing mechanism in order to change the small-scale statistics in
turbulent flows \cite{pandit,bif_prl,doering,vassilicos}. Typically,
for any given system, there exists a critical exponent, $b_c$,
characterizing the power law decaying of the forcing spectrum, $E(k)
\sim k^{-b}$, such that for $b < b_c$ the forcing is the leading
mechanism of energy exchange at all scales. In our case, the very
existence of Bolgiano scaling tells us that we fall in the latter
class. 
\begin{figure}
\begin{center}
\advance\leftskip-0.75cm
  \includegraphics[scale=0.7]{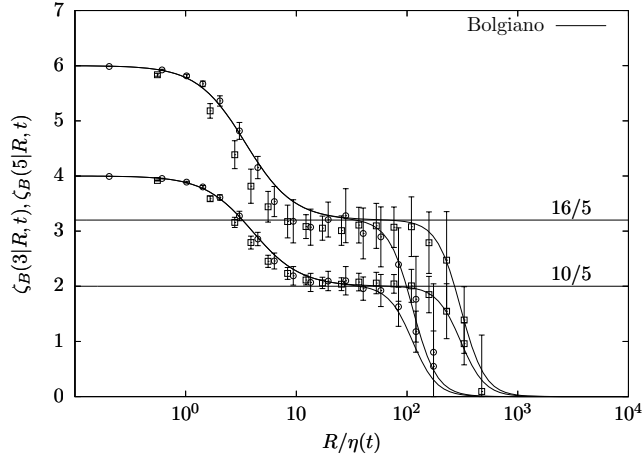}
  \caption{Local Scaling Exponent for buoyancy terms, $S_{B}^{(p)}(R,t)$, with $p=1,3$ for run (A) ($\circ$) and (C) ($\square$). The solid line corresponds to the dimensional estimate  (\ref{eq:batchelor2}) with $\zeta_B(p) = \zeta_\theta(1) + \zeta_{u_z}(p)$. The two horizontal lines give the expected Bolgiano scaling in the inertial range, $10/5$ ($p=3$)  and $16/5$ ($p=5$).  }
\label{fig:ls.fit.b3+5}
\end{center}
\end{figure}
The main interesting differences here, with previous
theoretical and numerical studies, is that the forcing mechanism is
also intermittent, i.e. very different from the typical
scaling-invariant Gaussian and delta-correlated in time power-law
forcing used in \cite{mazzino_rg,pandit,bif_prl}. Indeed, the high
intermittency of the temperature scaling shown in the previous
section, suggests the possibility that some degree of intermittency is
also hidden in the velocity field, even though a direct measure as the
one shown in figure \ref{fig:ls.fit.v4+6} rules out big effects. In
figure \ref{fig:ls.fit.b3+5} we are looking directly at the forcing
statistics entering in the equation of  high order velocity moments,
what we call the buoyancy structure functions in (\ref{eq:sf}),
$S_{B}^{(p)}(R,t)$. As one can see, even there it is hard to
disentangle any deviations from Bolgiano dimensional scaling. A
different scenario appears for the temperature flux structure
functions, $S_{F}^{(p)}(R,t)$, as defined in (\ref{eq:sf}), shown in
figure \ref{fig:ls.fitvt2}.  Here, a deviation from the dimensional
scaling is visible, due to the higher order of temperature fields with
respect to the velocity fields entering in these correlation
functions.
A possible way to highlight even better intermittent correction is to
look at the  behaviour of velocity and
temperature hyper-flatness:
\be
F_{u_z} (R,t)  = \frac{S_{u_z}^{(4)}(R,t)}{
  (S_{u_z}^{(2)}(R,t))^{2}}; \qquad
F_{\theta} (R,t)  = \frac{S_{\theta}^{(4)}(R,t)}{
  (S_{\theta}^{(2)}(R,t))^{2}}.
\ee
Any systematic dependence of flatness on the reference scale $R$ is
the signature of a non perfect self-similar statistics. Figure
\ref{fig:ess} shows temperature and velocity flatness at two different
times during the RT evolution. Temperature is clearly intermittent with
a flatness which increases at decreasing scale.  Velocity is more
noisy, nevertheless, our data cannot exclude a small scale-dependency
of flatness also for the latter, pointing towards small but detectable
breaking of self-similarity, i.e. corrections to the Bolgiano scaling
also for velocity. In the inset of the same figure, we show the
relative scaling of 4th and 6th order structure functions versus the
second order one, a procedure known as ESS in literature \cite{ess1,
  ess2}:
$$
S_{\theta}^{(p)}(R,t)\,\, vs\,\, S_{\theta}^{(2)}(R,t);\qquad
S_{u_z}^{(p)}(R,t)\,\, vs\,\, S_{u_z}^{(2)}(R,t);\qquad
$$
Here, a breaking of self similarity is detected as a deviation from
the dimensional scaling $S^{(p)}(R,t) =
(S^{(2)}(R,t))^{p/2}$. Deviations for the temperature/velocity are
strong/small and clearly detectable.

\begin{figure}
\begin{center}
\advance\leftskip-0.55cm
  \includegraphics[scale=0.7]{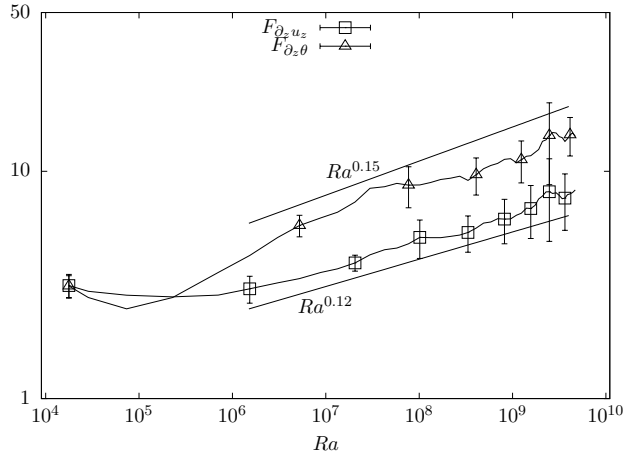}
  \caption{Flatness based on velocity $\square$, and temperature $\triangle$, gradients as a function of the Rayleigh number. Two power laws with the best fit for $Ra > 10^7$ are also shown as a guide for the eyes. }
\label{fig:flatness}
\end{center}
\end{figure}

 The above results suggest
that in order to highlight some possible non trivial scaling
properties in the velocity statistics, one needs to look at small
scales, where temperature intermittency becomes more intense and
possibly affects also the momentum equations. In 
  figure \ref{fig:flatness} we show the behaviour of the 
 flatness of velocity and
temperature derivatives during the RT evolution, i.e. at increasing
Rayleigh:
\be
F_{\partial_x u_z} (t)  = \frac{\langle  (\partial_x u_z)^p \rangle
}{\langle  (\partial_x u_z)^2 \rangle^{p/2}};\qquad
F_{\partial_x \theta} (t)  = \frac{\langle  (\partial_x \theta)^p \rangle }{\langle  (\partial_x \theta)^2 \rangle^{p/2}}.
\ee
Both small-scales temperature and velocity 
intermittency are increasing, with the temperature case much
faster. We fit a power law behaviour:
\be
F^{(p)}_{u_z} \sim Ra^{\xi_{\partial_x u_z}(p)}; \qquad
F^{(p)}_{\theta} \sim Ra^{\xi_{\partial_x \theta}(p)};  
\ee
with $\xi_{\partial_x u_z}(p) = 0.12 (5)$ and 
$\xi_{\partial_x \theta}(p) = 0.15 (5)$. While the result
for temperature does not  surprise, the result for velocity does,
supporting the existence of a small, but detectable intermittent
correction to the  2d Bolgiano scaling for the velocity field.

\section{Conclusions and further developments}
 
In this paper we have presented the results of a high resolution
numerical study of 2d Rayleigh-Taylor turbulence using a new thermal
lattice Boltzmann method.  The goal of the study was both
methodological and physical. Concerning the method, we validate and
assess the stability, accuracy and performances of the numerical
discrete kinetic algorithm used, showing that even when not perfectly
resolved at small scales, inertial and integral scale hydrodynamics is
well reproduced. This result opens the way to a systematic
exploitation of LBT algorithms also for fully developed turbulence.
Concerning the physics of RT turbulence in 2d, we have analyzed data
up to $Ra\sim 10^{11}$ and shown that the dynamics is dominated by a
Bolgiano phenomenology, i.e.  thermal fluctuations in the buoyancy
term are overwhelming the kinetic energy flux at all scales.  We have
also shown that:(i) a suitable Batchelor-like parametrization is able
to reproduce {\it scale-by-scale} the whole statistics at all scales,
over about 4 decades; (ii) temperature fluctuations show small-scales
intermittency, with scaling exponents tending to saturate at high
orders (see table \ref{tab:temp}), a signature of persistence of hot/cold fronts even at very
small scales \cite{fronts}; (iii) velocity statistics is much closer
to Bolgiano dimensional scaling even if small intermittent corrections
cannot be ruled out, especially concerning gradients evolution.

All these results are relevant for 3d thermal systems in presence of boundaries. Indeed, Bolgiano physics is believed to describe also thermal and velocity
fluctuations close to the boundary in real 3d convective
Rayleigh-B\`enard cells \cite{bolgiano3d-1, bolgiano3d-2}. The
existence of anomalous intermittent small scale fluctuations also in
these cases is relevant to control the physics of the viscous and
thermal boundary layers.

The algorithm presented here opens the way for natural generalization
to more complex situtations. First, it is trivially extendable to 3d
cases. Second, it can be further generalized including bulk forcing
terms in the internal energy equation, to describe reactive system
\cite{inprogress}. Third, it is under investigation the possibility to
couple the thermal LBT scheme with multi-component and/or multi-phase
LBT models \cite{SC1}, including non-trivial wettability properties at
the boundaries \cite{pre.nostro}: a case of interest to describe
convection of boiling systems.

We acknowledge useful discussion with A. Mazzino, G. Boffetta, A.  Celani and K. Sugiyama.
{We warmly thank the QPACE development team for support during the
implementation of our code and execution of our simulations. We furthermore
acknowledge access to QPACE and eQPACE during the bring-up phase of these
systems. Parts of the preliminary simulations were also performed on computing
resources made available by CASPUR under HPC Grant 2009. \label{sec:conclusions}


\clearpage

\clearpage

\end{document}